\documentclass{ifacconf}
\usepackage{amsmath}
\usepackage{amsfonts}
\usepackage{graphicx}      
\usepackage{natbib}        
\usepackage{xcolor}
\usepackage{framed}
\usepackage{multirow}
\usepackage{lineno}
\usepackage[utf8]{inputenc}
\usepackage{textcomp}
\usepackage{booktabs}
\usepackage{algorithm}
\usepackage{algpseudocode}
\usepackage{float}    
\makeatletter
\def\endfigure{\end@float}
\def\endtable{\end@float}
\makeatother

\let\ifacconfcaptionwidth\captionwidth
\usepackage[caption=false]{subfig}
\let\captionwidth\ifacconfcaptionwidth
\usepackage{soul}
\usepackage{color}
\usepackage{comment}
\usepackage{esvect}
\usepackage[normalem]{ulem}
\usepackage{silence}
\renewcommand{\S}{\mathcal{S}}
\newcommand{\X}{\mathcal{X}}
\newcommand{\B}{\mathcal{B}}
\newcommand{\G}{\mathcal{G}}
\newcommand{\D}{\mathcal{D}}
\allowdisplaybreaks

\begin{document}
\begin{frontmatter}

\title{\underline{E}psilon-Neighborhood \underline{D}ecision-Boundary \underline{G}overned \underline{E}stimation (EDGE) of 2D Black Box Classifier Functions\thanksref{support}}

\thanks[support]{This research was supported by the Ford Motor Company through the Ford-OSU Alliance Program.
DOI: 10.1016/j.ifacol.2025.12.274
}

\author[First]{M. Goutham},
\author[First]{R. DalferroNucci}, 
\author[First]{S. Stockar},
\author[Second]{M. Menon},
\author[Second]{S. Nayak},
\author[Second]{H. Zade},
\author[Second]{C. Patel},
\author[Second]{M. Santillo}

\address[First]{Department of Mechanical and Aerospace Engineering, The Ohio State University, 
   Columbus, OH 43210 USA (e-mail: goutham.1@osu.edu).}
\address[Second]{Ford Motor Company, Dearborn, MI 48109 USA.}

\begin{abstract}                
Accurately estimating decision boundaries in black box systems is critical when ensuring safety, quality, and feasibility in real-world applications. 
However, existing methods iteratively refine boundary estimates by sampling in regions of uncertainty, without providing guarantees on the closeness to the decision boundary and also result in unnecessary exploration that is especially disadvantageous when evaluations are costly. 
This paper presents $\varepsilon$-Neighborhood Decision-Boundary Governed Estimation (EDGE), a sample efficient and function-agnostic algorithm that leverages the intermediate value theorem to estimate the location of the decision boundary of a black box binary classifier within a user-specified $\varepsilon$-neighborhood.
To demonstrate applicability, a case study is presented of an electric grid stability problem with uncertain renewable power injection.
Evaluations are conducted on three test functions, where it is seen that the EDGE algorithm demonstrates superior sample efficiency and better boundary approximation than adaptive sampling techniques and grid-based searches.
\end{abstract}


\end{frontmatter}

\section{Introduction}
\vspace{0mm}

The successful deployment of real-world systems generally relies on satisfying safety, quality, or feasibility criteria, making it essential to precisely characterize how system parameters can vary before resulting in undesirable outcomes \citep{rogers2015feasibility}. 
This characterization involves identifying the decision boundary that separates ``desirable" and ``undesirable" system states \citep{feasibleDomain}.
However, in black box systems, explicit expressions describing the system are unavailable, making it impossible to analytically define this boundary, and thus, sampling methods are used \citep{karimi2019characterizing}. 

In safety critical applications, it is desirable to guarantee that the estimated boundary remains within a user-defined $\varepsilon$-neighborhood of the true decision boundary, where $\varepsilon$ is a parameter that specifies the permissible error in localization.
Black box boundary estimation methods typically refine function approximations by adaptively sampling in regions of high uncertainty to localize the boundary \citep{ono2024supclust,pepper2022adaptive}.

Existing methods lack the $\varepsilon$-neighborhood guarantee that is necessary to make design and operational decisions that rely on the estimated boundary.
Further, these techniques require hyperparameter tuning, demand high computational resources and their performance varies due to underlying randomness \citep{settles2009active}.
Due to the iterative nature of these methods, points far from the boundary are frequently explored, reducing sample efficiency, which is particularly detrimental when system state evaluations are computationally expensive.

To address these challenges, this paper presents the $\varepsilon$- Neighborhood Decision-Boundary Governed Estimation (EDGE), a novel algorithm that
\textcolor{black}{(i)} samples interior and exterior points strictly within a user-defined $\varepsilon$-neighborhood of the unknown decision boundary, maximizing sample efficiency, 
\textcolor{black}{(ii)} spans the decision boundary with minimal average surface distance, a metric that quantifies total approximation error between the estimated and true boundary \citep{heimann2009comparison}, and
\textcolor{black}{(iii)} requires no hyperparameter tuning, making it applicable across various black box system characteristics.

\textcolor{black}{
EDGE is initialized by applying the intermediate value theorem (IVT) to localize the decision boundary within an $\varepsilon$-neighborhood, under the assumption that a continuous boundary exists in the domain being explored.
Next, EDGE sequentially queries the intersections of $\varepsilon$-radius circles centered at interior and exterior points to trace the boundary with high sample efficiency and localization accuracy.
However, the use of circle intersections restricts EDGE to two-dimensional decision boundary localization.}

After defining the problem and the EDGE algorithm, this paper demonstrates its effectiveness through a case study involving grid stability, assessing feasible ranges of power injection from renewable energy sources.
Next, the performance of the EDGE algorithm is evaluated using three highly nonlinear two dimensional benchmark functions.
Its computational efficiency and accuracy are then compared against a na\"ive $\varepsilon$-spaced grid sampling method that provides $\varepsilon$-neighborhood guarantees, and an active learning method that uses support vector machines.


\section{Problem Description}\label{sec:problem_formulation}
\begin{figure}[b]
    \centering
  \subfloat[The $\varepsilon$-neighborhood of $\partial S$\label{im:BoundaryDefinition} ]{%
       \includegraphics[trim =0mm 100mm 245mm 0mm, clip, width=0.49\linewidth]{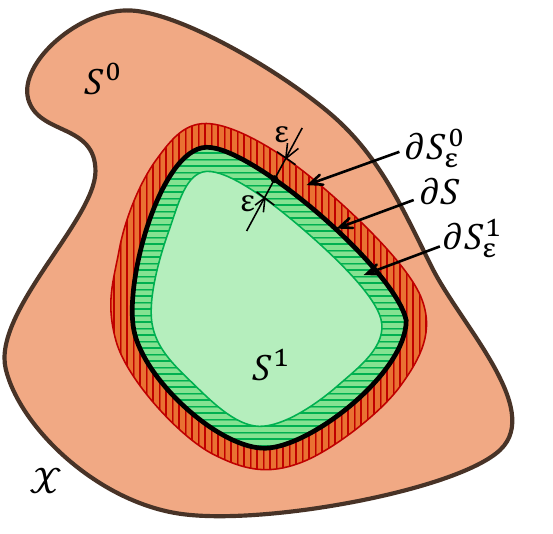}}
       \vspace{0mm}
\subfloat[Sampling in $\partial S^{0}_\varepsilon$ and $\partial S^{0}_\varepsilon$ \label{im:distanceFromBoundary}]{%
       \includegraphics[trim =100mm 105mm 150mm 0mm, clip, width=0.49\linewidth]{images/Fig_1ab.pdf}}
  \caption{Definition of sets}
  \label{IM:setDefinition} 
  \vspace{0mm}
\end{figure}

Let $\X \subseteq \mathbb{R}^2$ be a compact domain of interest. 
Consider a black box boolean-valued classifier function $f(\cdot) : \mathcal{X} \to \mathbb{B}$. 
The value $f(x) = 1$ may imply some desirable property being studied, such as $x$ being classified as \emph{safe} in some engineering operation, or as \emph{feasible} in a design or optimization problem, or that $x$ \emph{passes} some defined performance or quality test, etc.
The \emph{absence} of the indicated property corresponds to $f(x) = 0$ in this binary representation.

Let the set $S^1 := \{ x \in \mathcal{X} : f(x) = 1 \}$ denote the set of all \textit{interior points} in $\X$ that are desirable.
Denote its complementary set of \textit{exterior points} by $S^0 := \{ x \in \mathcal{X} : f(x) = 0 \}$.
The \textit{decision boundary} separating the two sets is given by $\partial S$ as indicated in Fig. \ref{im:BoundaryDefinition}, and corresponds to the interface where $f(x)$ changes in value. 

Defining $\partial S$ is generally infeasible, especially for a black box function due to a lack of access to its closed-form expression.
To overcome this, a tolerance parameter $\varepsilon$ is introduced to define two bounding sets that capture $\partial S$: 
\begin{subequations}
\label{eq:nbhd}  
\begin{align}
&\textrm{The inner $\varepsilon$-neighborhood set of $\partial S$:} \notag\\
\label{eq:nbhd1}
     &\quad \quad  \partial S^{1}_\varepsilon := 
    \Bigl\{\, x \in \mathcal{X} : f(x) = 1,\; \mathrm{d}\bigl(x, \partial S\bigr) \le \varepsilon \Bigr\}\\
&\textrm{The outer $\varepsilon$-neighborhood set of $\partial S$:} \notag\\
\label{eq:nbhd0}    
     &\quad \quad \partial S^{0}_\varepsilon:=
        \Bigl\{\, x \in \mathcal{X} : f(x) = 0,\; \mathrm{d}\bigl(x, \partial S\bigr) \le \varepsilon \Bigr\}
\end{align}
\end{subequations}
 In Eq.(\ref{eq:nbhd}), the distance from a point $x$ to $\partial S$ is given by $\mathrm{d}\bigl(x, \partial S\bigr) := \inf_{y \in \partial S} \|x-y\|$.
The striped regions in Fig.\ref{IM:setDefinition} depict $\partial S^{1}_\varepsilon$ and $\partial S^{0}_\varepsilon$, and comprise the regions in $\X$ within an $\varepsilon$-distance from decision boundary $\partial S$, i.e., the $\varepsilon$-neighborhood of $\partial S$.
The parameter $\varepsilon$ can be chosen based on the desired localization accuracy of the decision boundary, whereby a small $\varepsilon$-value indicates a need for a high degree of certainty in the localization of $\partial S$.

\textcolor{black}{
The $\varepsilon$-neighborhood also accommodates stochasticity in function evaluations, where repeated queries at the same point $x$ yield different $f(x)$ outputs, making it impractical to localize a well-defined decision boundary.
In this context, the $\varepsilon$-neighborhood provides a soft transition between the $S^1$ and $S^0$ regions, provided that the value of $\varepsilon$ is chosen to absorb the expected variability.
}

Despite the $\varepsilon$-tolerance, defining $\partial S^{1}_\varepsilon$ and $\partial S^{0}_\varepsilon$ as per Eqs.~(\ref{eq:nbhd1}) and (\ref{eq:nbhd0}) remains impossible without explicit knowledge of $\partial S$. 
To address this, we propose a sampling-based approach that constructs inner and outer approximating point sets, $\partial \tilde{S}^1 \subset \partial S^{1}_\varepsilon$ and $\partial \tilde{S}^0 \subset \partial S^{0}_\varepsilon$, which bracket the true boundary within its $\varepsilon$-neighborhood.

Sample efficiency is an important consideration when attempting to approximate $\partial S$ through such a sampling-based approach.
Minimizing the number of function queries is necessary when queries to $f(\cdot)$ are expensive, involving high-fidelity simulations, large-scale combinatorial optimization problems, or physical experiments.

The objectives when estimating $\partial S$ are:
\vspace{-0.5mm}
\begin{enumerate}
    \item \textit{Sample Efficiency:} Minimize the number of queries needed to identify points in $\partial \tilde{S}^{0}$ and $\partial \tilde{S}^{1}$, thereby approximating $\partial S$ within its $\varepsilon$-neighborhood.
    \item \textit{Approximation Accuracy:} Minimize the average surface distance (ASD) between the decision boundary $\partial S$,  $\partial \tilde{S}^{0}$ and $\partial \tilde{S}^{1}$, given by:
\end{enumerate}
\vspace{-1.5mm}
\begin{subequations}
\begin{align}
\label{eq: 2sided ASSD}  
\quad \quad 
d(\partial \tilde{S}^0,\partial S,\partial \tilde{S}^1) 
    &= d(\partial \tilde{S}^0,\partial S) + d(\partial \tilde{S}^1,\partial S) \\
\label{eq: ASSD}
\textrm{where, } d(\partial \tilde{S},\partial S) 
    &= \frac{1}{|\partial \tilde{S}|+|\partial S|} \Bigg[
    \sum_{x\in \partial \tilde{S}}d(x,\partial S) \notag\\
    &\quad \quad \quad \quad  \quad \quad+ \sum_{y\in \partial S}d(y,\partial \tilde{S})
    \Bigg]
\end{align}
\end{subequations}
Minimizing $d(\partial \tilde{S}^0,\partial S,\partial \tilde{S}^1)$ ensures that points in $\partial \tilde{S}^0$ and $\partial \tilde{S}^1$ adequately span the boundary $\partial S$, avoiding the trivial solution of finding a single point close to $\partial S$.
Some points of $\partial \tilde{S}^1$ and $\partial \tilde{S}^0$ and are shown in illustrative Fig. \ref{im:distanceFromBoundary}.
Although $\partial S$ is unknown, for evaluation purposes, it can be closely approximated by choosing a sufficiently low $\varepsilon$. 

\section{Proposed EDGE Algorithm}


The IVT is the key enabler of the EDGE algorithm, ensuring that sampled points lie in the $\varepsilon$-neighborhood of $\partial S$.
If there exist points $x^1_n$ and $x^0_m$ satisfying $f(x^1_n)=1$ and $f(x^0_m)=0$, and the distance $\|x^1_n  - x^0_m\| \leq \varepsilon$, then the IVT when applied in the boolean context guarantees that $\partial S$ intersects the segment connecting $x^1_n$ and $x^0_m$.
This means that necessarily, both $x^1_n\in \partial \tilde{S}^{1}$ and $x^0_m\in \partial \tilde{S}^{0}$ because there exists a transition from function value $1$ to $0$ somewhere along the line segment joining these two points, and therefore, $d(x^1_n,\partial S)\leq \varepsilon$ and $d(x^0_m,\partial S)\leq \varepsilon$. 

In practice, a boundary $\partial S$ need not exist for $f(\cdot)$ in the domain $\X$, but the EDGE algorithm can only be initialized if at least one interior and one exterior point are found.
If multiple queries are needed to find these points, denote the closest pair of points as $\overline{x}^{in}_0 \in S^1$ and $\overline{x}^{out}_0 \in S^0$. 

\begin{figure}[t]
\vspace{0mm}
    \centering
        \includegraphics[trim =0mm 135mm 267mm 0mm, clip, width=0.5\linewidth]{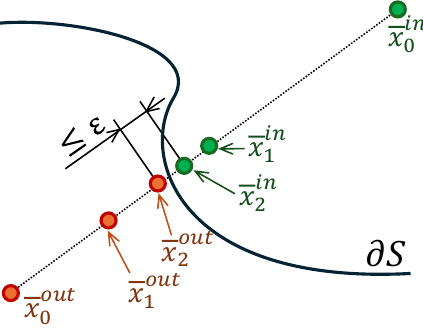}
        \vspace{0mm}
  \caption{Bisection to $\varepsilon$ ball of $\partial S$}
  \label{im:bisection} 
\end{figure}

The bisection procedure is first used to localize $\partial S$ along the segment connecting $\overline{x}^{in}_0$ and $\overline{x}^{out}_0$, as described by Alg.~\ref{alg:bisection}. 
If the point at the midpoint of $\overline{x}^{in}_0$ and $\overline{x}^{out}_0$ has a function value of $1$, it is appended to the ordered set $\overline{x}^{in}$ as $\overline{x}^{in}_1$. 
This is as shown in the illustrative example of Fig. \ref {im:bisection}, repeatedly querying the midpoint of the most recently appended points of sets $\overline{x}^{in}$ and $\overline{x}^{out}$, until the distance between the last appended points of $\overline{x}^{in}_0$ and $\overline{x}^{out}_0$ is less than $\varepsilon$, thus converging to the $\varepsilon$-neighborhood of $\partial S$.


\begin{figure}[b]
\vspace{0mm}
    \centering
        \includegraphics[trim =72mm 135mm 206mm 0.5mm, clip, width=0.65\linewidth]{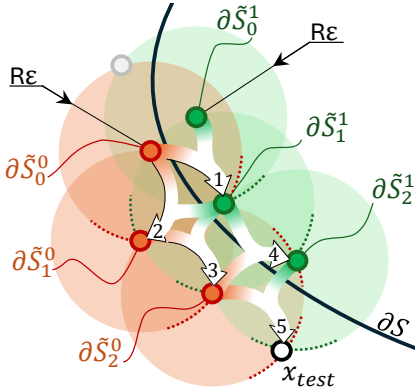}
        \vspace{-4mm}
  \caption{$\varepsilon$-stepping along $\partial S$ boundary}
  \label{im:decisionBoundaryWalk}
\end{figure}

\begin{algorithm}[t]
\caption{Bisection to the $\varepsilon$-neighborhood of $\partial S$}
\label{alg:bisection}
\begin{algorithmic}[1]
\Function{Bisect}{$\overline{x}^{in}_{0}, \overline{x}^{out}_{0}, \varepsilon$}
    \State \label{algline:init-out}
    $\overline{x}^{in} \gets \overline{x}^{in}_{0}; 
    \quad 
     \overline{x}^{out} \gets \overline{x}^{out}_{0}$

    \While{\label{algline:while}
    $\| \overline{x}^{in}_{end} - \overline{x}^{out}_{end}\|\geq \varepsilon$}
        \State \label{algline:testpoint}
        $x_\text{test} 
        \gets
        (\overline{x}^{in}_{end} + \overline{x}^{out}_{end})/2$

        \If{$f(x_\text{test}) = 0$}    \Comment{Exterior test point}
            \label{algline:feascheck}
            
            \State Append $x_{\text{test}}$ to $\overline{x}^{out}$
            \label{algline:assign-out}
            
        \Else    \Comment{Interior test point}
            \State Append $x_{\text{test}}$ to $\overline{x}^{in}$
            \label{algline:assign-in}
            
        \EndIf
    \EndWhile
    
    \State \Return $\overline{x}^{in}, \overline{x}^{out}$
    \label{algline:return}
\EndFunction
\end{algorithmic}
\end{algorithm}

\begin{algorithm}[t]
\caption{Decision Boundary $(\partial S)$ Walk}
\label{alg:decisionStepping}
\begin{algorithmic}[1]
\Function{DecisionBoundary}{$\overline{x}^{in}, \overline{x}^{out}, \partial \mathcal{X}, \varepsilon$}
    \State $\partial \tilde{S}^1 \gets \overline{x}^{in}_{end}$;
        \quad 
        $\partial \tilde{S}^0 \gets \overline{x}^{out}_{end}$
\Repeat
    \State \label{alg:circleIntersection}
    $C 
        \gets 
        \{\textrm{circle}(\partial \tilde{S}^1_{end},\varepsilon) 
        \cap 
        \textrm{circle}(\partial \tilde{S}^0_{end},\varepsilon)\}$ 
        
    \State  \label{alg:circleIntersectiontest}
        $x_{\text{test}}
        \gets
        x\in C|(\partial \tilde{S}^0_{end}-\partial \tilde{S}^1_{end})\boldsymbol{\times} (x-\partial \tilde{S}^1_{end}) >0 $
            \If {$x_{test} \in \mathcal{X}$}
                    \If{$f(x_\text{test}) = 0$}    \Comment{Exterior test point}
            \label{algDecision:feascheck}
            
            \State Append $x_{\text{test}}$ to $\overline{x}^{out}$
            \label{algDecision:assign-out}
                    
                \Else    \Comment{Interior test point}
                    \State Append $x_{\text{test}}$ to $\overline{x}^{in}$
                    \label{algDecision:assign-in}
                    
                \EndIf
            \Else
                \State \label{alg:outsideDomain}$\textsc{domainBoundary}(\partial \tilde{S}^1, \partial \tilde{S}^0, \partial \mathcal{X}, \varepsilon)$
            \EndIf
            \Until{
    $\min \{ \|x_{\text{test}}-\partial \tilde{S}^1_0 \|,\|x_{\text{test}}-\partial \tilde{S}^0_0 \| \}\leq \varepsilon$\\
    $ \quad \quad ~~ \textrm{and} \min \{|\partial \tilde{S}^1|,|\partial \tilde{S}^0|\} >1$
}
    \State \Return $\partial \tilde{S}^1, \partial \tilde{S}^0$
\EndFunction
\end{algorithmic}
\end{algorithm}

The last appended points of the interior bisection sequences $\overline{x}^{in}_{end}$ and $\overline{x}^{out}_{end}$ are the first points $\partial \tilde{S}^1_0$ and $\partial \tilde{S}^0_0$ of the inner and outer point sets respectively.
These points initiate the decision boundary walk of Alg. \ref{alg:decisionStepping}.

Let $\textrm{circle}(x_c,\varepsilon)$ represent the $\varepsilon$-radius circle centered at $x_c$, or equivalently, the set $\{x: \|x-x_c\| = \varepsilon \}$.
As shown in Fig. \ref{im:decisionBoundaryWalk}, the intersection of $\textrm{circle}(\partial \tilde{S}^1_0,\varepsilon)$ and $\textrm{circle}(\partial \tilde{S}^0_0,\varepsilon)$ is first obtained.
This produces two intersection points that form set $C$ of line \ref{alg:circleIntersection} in Alg. \ref{alg:decisionStepping}.
The intersection point which has a positive cross product relative to the points $\partial \tilde{S}^1_0$ and $\partial \tilde{S}^0_0$ is selected as $x_{test}$, per line \ref{alg:circleIntersectiontest} of Alg. \ref{alg:decisionStepping}.
The selection of a positive cross product is arbitrary, and could instead be negative, so long as the direction sense is maintained throughout Alg. \ref{alg:decisionStepping}.

In the illustrative example of Fig. \ref{im:decisionBoundaryWalk}, the first $x_{test}$ point returned $f(x_{test})=1$, and is appended as $\partial \tilde{S}^1_1$ of inner set $\partial\tilde{S}^1$.
Next, the intersection points of $\textrm{circle}(\partial \tilde{S}^1_1,\varepsilon)$ and $\textrm{circle}(\partial \tilde{S}^0_0,\varepsilon)$, the last point of the $\tilde{S}^0$ set are obtained.
The point that satisfies a positive relative cross product is queried, and appended to the appropriate set.
In this manner, Alg. \ref{alg:decisionStepping} repeatedly steps by an $\varepsilon$ distance along both sides of $\partial S$ in its $\varepsilon$-neighborhood.

\begin{figure}[b]
\vspace{0mm}
    \centering
        \includegraphics[trim =0mm 150mm 235mm 0mm, clip, width=0.95\linewidth]{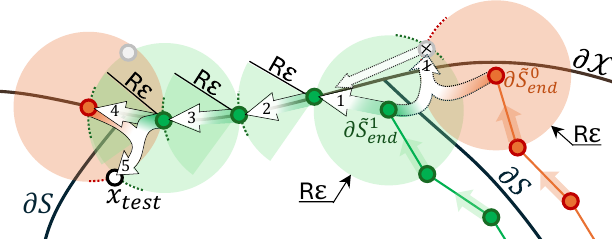}
        \vspace{-2mm}
  \caption{$\varepsilon$-stepping along $\partial \X$ boundary}
  \label{IM:domainBoundary} 
\end{figure}

If a test point $x_{test}$ is found to be outside the domain $\X$, an $\varepsilon$-stepping walk is conducted along $\partial \X$, as detailed in Alg.~\ref{alg:domainBoundaryWalk} and illustrated by Fig. \ref{IM:domainBoundary}.
First, two candidate points $x_1,x_2$ on $\partial \mathcal{X}$ are identified that are $\varepsilon$ away from the $\partial \tilde{S}^1_{end}$ point.
As defined in step \ref{algBS:chooseTestPoint} of Alg.~\ref{alg:domainBoundaryWalk}, the point farther away from the $\partial \tilde{S}^0_{end}$ point is selected as $x_{test}$.
This preference ensures that the boundary intersecting $\partial \mathcal{X}$ is identified. 

The while-loop of steps \ref{algBS:whileStart}-\ref{algBS:whileEnd} continues to select test points along $\partial \X$ in the same direction sense, appending encountered interior points to $\partial \tilde{S}^1$.
This is executed until an exterior point is detected, which is appended to $\partial \tilde{S}^0$, and indicates that $\partial S$ has been detected, re-initiating Alg. \ref{alg:decisionStepping}.

\begin{algorithm}[b]
\caption{Domain Boundary $(\partial \X)$ Walk}
\label{alg:domainBoundaryWalk}
\begin{algorithmic}[1]
\Function{domainBoundary}{$\partial \tilde{S}^1, \partial \tilde{S}^0, \partial \mathcal{X}, \varepsilon$}
    \State \label{algBS:twoPoints}
        $\{x_1,x_2\} 
        \gets 
        \{ x\in \partial \mathcal{X}  
        \mid
        \|x-\partial \tilde{S}^1_{end}\|=\varepsilon\}$ 
    \State \label{algBS:chooseTestPoint}
        $x_{test} 
        \gets 
        \arg\max_{x\in \{x_1,x_2\}} 
        \|x-\partial \tilde{S}^0_{end}\|$
    \While{$f(x_\text{test}) = 1$} \label{algBS:whileStart}       

        \State \label{algBS:updateIn}
            Append $x_{\text{test}}$ to $\partial \tilde{S}^1$

        \State \label{algBS:recomputePoints}
        $x_{test} 
        \gets 
        \{ x\in \partial \X \setminus \partial \tilde{S}^1  
        \land
        \|x-\partial \tilde{S}^1_{end}\| = \varepsilon\}$       
        
    \EndWhile \label{algBS:whileEnd}
    
    \State \label{algBS:updateOut}
    Append $x_{\text{test}}$ to $\partial \tilde{S}^0$

    \State \label{algBS:return}
        \Return $\partial \tilde{S}^1, \partial \tilde{S}^0$
\EndFunction
\end{algorithmic}
\end{algorithm}

The decision boundary walk of Alg. \ref{alg:decisionStepping} is continued until a test point is within an $\varepsilon$ distance away from one of the initializing points $\partial \tilde{S}^1_0$ or $\partial \tilde{S}^0_0$.
This indicates that the entire length of $\partial S \in \X$ has been spanned within an $\epsilon$-tolerance and the EDGE algorithm is terminated, returning the $\partial \tilde{S}^1$ and $\partial \tilde{S}^0$ sets, i.e., the inner and outer approximation sets of $\partial S\in \X$.

The EDGE algorithm is function agnostic since it has no hyper-parameters that require tuning, and can be applied to any black box function.
While the bisection algorithm produces unavoidable function queries in regions outside $\partial \tilde{S}^1$ and $\partial \tilde{S}^0$, the $\varepsilon$-stepping algorithms for the decision and domain boundaries yield a 100\% sample efficiency, whereby every sampled point is appended to $\partial \tilde{S}^1$ or $\partial \tilde{S}^0$.

\section{Computational Experiments}


\subsection{Case Study: Power Grid with Renewable Sources}
Consider a grid network represented by a set of buses $i\in\B$, generators $g\in\G$ and transmission lines $ij\in\mathcal{L}$.
The DC Optimal Power Flow (DC-OPF) formulation is a grid model commonly used for stability analysis, enabling real-time response to varying demands and renewable power generation ~\citep{purchala2005usefulness}.
The minimization problem of Eq. ~\eqref{eq:dcOPFsubset} determines the most economical operating point for a power grid, where $c_g$ is the operational cost associated with generator $g$ and $p_g$ is its power output:
\begin{subequations}  \label{eq:dcOPFsubset}
\allowdisplaybreaks
\begin{align}   
\label{eq:ProbForm}
J  = \min_p \sum_{\substack{g\in \mathcal{G}}} c_g p_{g} 
\quad \quad \quad \quad &\\
\label{eq:nodal}
\textrm{s.t.} \quad 
\sum_{g\in \G_i} p_{g} + 
\sum_{ji\in \mathcal{L}} p_{ji} = \sum_{d\in \D_i} p_{d},
    \quad \forall i\in \B &
\\
\label{eq:genLimits}
\underline{p}_g \leq p_{g}\leq \overline{p}_g ,
\quad \forall g\in \G &
\\
\label{eq:tranLimits}   
-\overline{p}_{ij} \leq p_{ij} \leq \overline{p}_{ij},
\quad \forall ij\in \mathcal{L} &
\\ 
\label{eq:phaseDiff} 
p_{ij} = -b_{ij}(\theta_{i}-\theta_{j}),
\quad \forall ij\in \mathcal{L} &
\\ 
\label{eq:phaseDiffLimits}  
\underline{\theta}_{ij} \leq \theta_{i} - \theta_{j} \leq \overline{\theta}_{ij}, \: \forall ij \in \mathcal{L} &
\end{align} 
\end{subequations}
The constraints defined by Eq.~\eqref{eq:nodal} - \eqref{eq:phaseDiffLimits} govern grid power flow. 
Nodal power balance at each bus $i \in \mathcal{B}$ is imposed by Eq. \eqref{eq:nodal}, ensuring that the total generation $\sum_{g\in \G_i} p_{g}$ and incoming line flow $\sum_{ji\in \mathcal{L}} p_{ji}$ are equal to the total demand $\sum_{d\in \D_i} p_{d}$ at bus $i$.
The power generated at $g \in \G$ and transmitted through line $ij \in \mathcal{L}$ are constrained by Eq.\eqref{eq:genLimits} and Eq.\eqref{eq:tranLimits} to be within limits defined by $[\underline{p}_g, \overline{p}_g]$ and $[-\overline{p}_{ij}, \overline{p}_{ij}]$ respectively.
The power transmitted through each line $p_{ij}$ is defined in Eq.~\eqref{eq:phaseDiff} using line susceptance $b_{ij}$ and the difference in voltage phase angles $\theta_i$ and $\theta_j$.
The permissible phase angle differences between buses $i,j$ is defined in Eq.~\eqref{eq:phaseDiffLimits} to be within $[\underline{\theta}_{ij},\overline{\theta}_{ij}]$.

Renewable energy sources, such as solar arrays and wind farms, produce power outputs governed by changing environmental conditions that are not directly controllable.
Their power output therefore appears as system parameters that vary as a function of incident solar irradiance or wind speed respectively. 
When connected to the grid, the variations in renewable power injection can disrupt nodal power balance or cause violations of constraints defined by Eq. ~\eqref{eq:nodal} - \eqref{eq:phaseDiffLimits}.
Consequently, the informed incorporation of renewable power sources requires the careful selection of power ratings of these renewable sources to maintain stable operation despite their inherent variability.

For this illustrative case study, the widely utilized IEEE 5-bus system and its defined parameters are used \citep{li2010small}.
Generators 1 and 5 are replaced by renewable power sources rated at 10 and 7 MW, respectively. 
The domain $\X$ is therefore defined as $[0,10]\times[0,7]$, reflecting the expected variation in power output for these two sources. 

\begin{figure}[t]
\vspace{-2mm}
    \centering
        \includegraphics[trim =63mm 104mm 68mm 110mm, clip, width=0.75\linewidth]{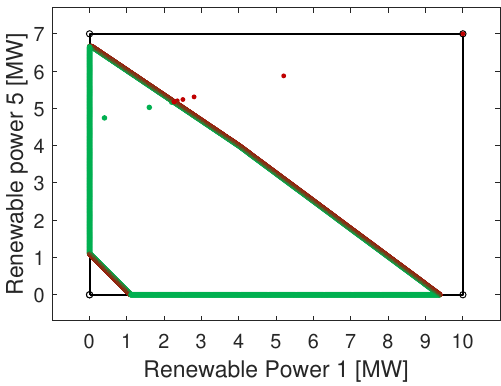}
        \vspace{-1mm}
  \caption{Power injection $\partial S$ ($\varepsilon = 0.01 MW$)}
  \label{IM:powergrid} 
\end{figure}

The first interior point, $\overline{x}^{in}_{0}$, is selected as the nominal grid operating state prior to the integration of renewable sources, marked in green at the coordinate $(0.4, 4.74)$ in Fig. \ref{IM:powergrid}.
To identify an exterior point, the four corner points are then evaluated. 
In this case, the first corner inspected, $(10, 7)$ is confirmed to be $\overline{x}^{out}_{0}$, an exterior point because it violated one of the constraints defined in Eqs. \eqref{eq:nodal}–\eqref{eq:phaseDiffLimits}.
The bisection algorithm is initiated with $\overline{x}^{in}_{0}$ and $\overline{x}^{out}_{0}$, converging to the specified $\varepsilon$-neighborhood, after which the $\varepsilon$-stepping algorithms are used to span $\partial S \in \X$.
When the $\varepsilon$ value was set to $0.01$ MW, the true boundary was delineated by sampling $2,866$ points, while an $\varepsilon$ value of $0.1$ MW required $282$ points. 

This illustrative example treated the DC-OPF feasibility problem as a black box classifier $f(\cdot)$, i.e., it remained inaccessible to the EDGE algorithm.
Despite this, the resulting approximated boundary is clearly a polygon, as expected for this linear formulation.

\subsection{Nonlinear Test Functions}

For the following test cases, the boundary $\partial S$ corresponds to the level set where a 2D test function attains a specified constant value.
For the Rosenbrock test function defined by Eq. \eqref{eq:rosenbrock}, the domain is set to $\X_f:=[-6,6]\times[-2,10]$, and $\partial S$ is defined by $f(x,y) = 100$.
Interior points satisfy $f(x,y) < 100$ and exterior points satisfy $f(x,y) > 100$.
\begin{subequations}
\label{eq:testFx}
\begin{align}
\label{eq:rosenbrock}
    f(x, y) = &(1 - x)^2 + 100 (y - x^2)^2 
    \\
\label{eq:price}
    g(x, y) = &[1 + (x + y + 1)^2 \nonumber \\
     \times &(19 - 14x + 3x^2 - 14y + 6xy + 3y^2) ] \nonumber \\
     \times &[30 + (2x - 3y)^2 \nonumber \\
    \times&(18 - 32x + 12x^2 + 48y - 36xy + 27y^2) ]  \\
\label{eq:beale}
    h(x, y) = &(1.5 - x + xy)^2 + (2.25 - x + xy^2)^2 \nonumber \\
     &+ (2.625 - x + xy^3)^2 
\end{align}
\end{subequations}
The Goldenstein-Price and Beale test functions defined by Eq. \eqref{eq:price} and \eqref{eq:beale} are similarly binarized, setting $g(x,y) = 100$ and $h(x,y) = 3e5$ to define decision boundaries respectively. 
Their domains are defined by $\X_g:=[-8,10]\times[-7,6]$ and $\X_h:=[-6,6]\times[-7,7]$ respectively.

\subsection{Benchmark Alg. 1: Na\"ive Grid Sampling}

\begin{figure}[t]
    \centering
      \vspace{0mm}
      \subfloat[Na\"ive localization \label{im:Goldenstein_GS}]{%
      \includegraphics[trim =60mm 105mm 55mm 102mm, clip, width=0.47\linewidth]{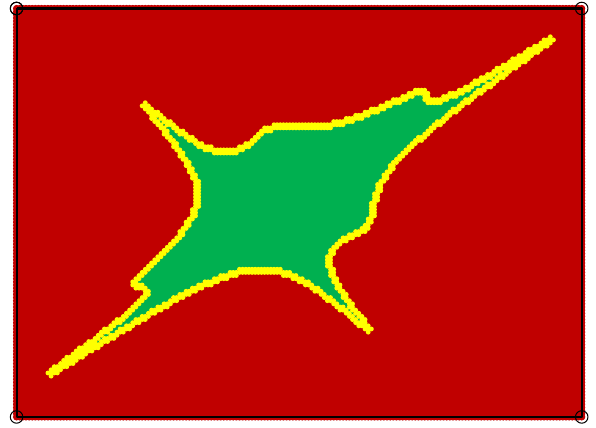}}
  \hspace{5mm}
\subfloat[EDGE localization \label{im:Goldenstein_ENSDB} ]{%
       \includegraphics[trim =84mm 122mm 81mm 120mm, clip, width=0.46\linewidth]{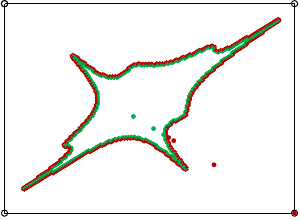}}
       \caption{Goldenstein-Price test function, $g(x,y) = 100$}
  \label{IM:goldenstein} 
  \vspace{0mm}
\end{figure}
Due to a lack of methods in literature that localize $\partial S$ to within its $\varepsilon$-neighborhood, in this paper, a simple algorithm is developed as a benchmark that meets the outlined objectives.
This na\"ive approach samples the domain $\mathcal{X}$ over a uniform grid of $\varepsilon$-spaced points, querying each grid point to determine if it belongs to $S^1$ or $S^0$.
After every grid point has been classified, point labels are inspected in each $\varepsilon$-neighborhood, detecting $\partial \tilde{S}^0$ and $\partial \tilde{S}^1$ points based on the encountered label transitions.
This guarantees $\varepsilon$-neighborhood localization of $\partial S$ while also spanning the entire boundary, minimizing ASD, i.e., $d(\partial \tilde{S}^0,\partial S,\partial \tilde{S}^1)$.

As illustrated in Fig. \ref{im:Goldenstein_GS}, this approach tests every point in the grid to localize $\partial S$, \textcolor{black}{marked by yellow points.}
Compared with the EDGE algorithm whose sampled points are shown in Fig. \ref{im:Goldenstein_ENSDB}, the na\"ive approach has poor sample efficiency and is computationally prohibitive for large domains.
Table \ref{tab:perfComp} lists outcomes for the 3 test functions.

\begin{table*}[ht]
\centering
\caption{Performance comparison of 3 decision boundary estimating methods}
\label{tab:perfComp}
\begin{tabular}{rc|ccc|ccc|ccc}
\hline
\multicolumn{1}{c}{\textbf{}}              & \textbf{}           
& \multicolumn{3}{c|}{\textbf{ASD}}                           
& \multicolumn{3}{c|}{\textbf{Sample Count}}   
& \multicolumn{3}{c}{\textbf{Computation Time {[}s{]}}}       
\\
\multicolumn{1}{c}{\textbf{Test Function}} 
& \textbf{$\varepsilon$} 
& \textbf{EDGE} 
& \textbf{ALB $\mu (\sigma)$} 
& \textbf{Grid} 
& \textbf{EDGE} 
& \textbf{ALB} 
& \textbf{Grid} 
& \textbf{EDGE} 
& \textbf{ALB $\mu (\sigma)$} 
& \textbf{Grid} 
\\ \hline
Rosenbrock      & 0.1                 
& 0.103         & 0.845~(0.665)          & 0.112         
& 973           & $\leftarrow$          & 14641         
& 0.08          & 44.7~(14.060)& 0.60          
\\
& 0.05
& 0.052         & 0.600~(0.520)         & 0.051         
& 1948          & $\leftarrow$          & 58081         
& 0.13          & 1378.23~(356.82)     & 6.69          
\\ \hline
Goldenstein-Price  & 0.1                 
& 0.112         & 0.661~(0.247)         & 0.111         
& 1101          & $\leftarrow$          & 23711         
& 0.06          & 54.11~(16.62)        & 1.15          
\\
& 0.05 
& 0.052         & 0.530~(0.170)          & 0.051         
& 2262         & $\leftarrow$           & 94221        
& 0.14         & 1184.40~(207.62)              & 21.22       
\\ \hline
Beale          & 0.1                 
& 0.116        & 0.515~(0.150)          & 0.112         
& 909          & $\leftarrow$           & 17061         
& 0.07         & 18.07~(3.09)          & 0.69          
\\
& 0.05         
& 0.055        & 0.374~(0.123)          & 0.054        
& 1903         & $\leftarrow$           & 67721         
& 0.13         & 764.52~(175.59)       & 7.88          
\\ \hline
\end{tabular}
\end{table*}

\subsection{Benchmark Alg. 2: Active Learning at the Boundaries}


The performance of the EDGE algorithm is also compared with an active learning at the boundaries (ALB) algorithm which uses an iterative Support Vector Machine (SVM) approach \citep{kremer2014active}.
SVM is a supervised machine learning method that seeks to identify the decision boundary that separates two labeled sets of data points with the largest possible margin \citep{vapnik1995support}.  

SVMs use kernel functions $k(x_i,x_j)$ to quantify the correlation between labels at points $x_i$ and $x_j$.
In black box applications, the Radial Basis Function (RBF) kernel is typically chosen \citep{albahar2021robust}.
This kernel is defined by $k(x_i,x_j)=exp(-\| x_i-x_j\|^2/2\sigma^2)$, where $\sigma$ is a width hyperparameter that determines the influence of each training example on the predicted boundaries.

The active learning component iteratively samples from regions of high uncertainty, i.e., the predicted decision boundary region. It executes the following:
\begin{enumerate}
    \item Randomly generate $n_{init}$ unlabeled points in $\X$ 
    \item Label them using $f(\cdot): \X \to \{0,1\}$
    \item Train SVM using labeled set available
    \item Identify $n_{learn}$ points at the predicted boundary
    \item Repeat steps 2 to 4 until budget is exhausted
\end{enumerate}

\begin{figure}[t]
    \centering
      \vspace{-1mm}
\subfloat[1\textsuperscript{st} ALB iteration \label{im:Beale1} ]{%
       \includegraphics[trim =92mm 123mm 90mm 117mm, clip, width=0.34\linewidth]{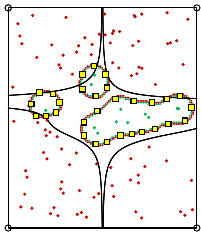}}
       \hspace{5mm}
\subfloat[15\textsuperscript{th} ALB iteration \label{im:Beale5}]{%
       \includegraphics[trim =92mm 123mm 90mm 117mm, clip, width=0.34\linewidth]{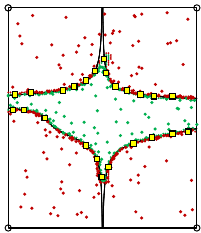}}\\
\subfloat[Final ALB iteration \label{im:BealeEnd} ]{%
       \includegraphics[trim =92mm 123mm 90mm 117mm, clip, width=0.34\linewidth]{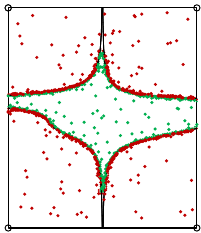}}
       \hspace{5mm}
\subfloat[EDGE localization \label{im:BealeEDGE}]{%
       \includegraphics[trim =92mm 123mm 90mm 117mm, clip, width=0.34\linewidth]{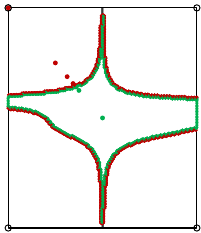}}
  \caption{Beale Test Function, $h(x,y)=3e5$}
  \label{IM:Beale} 
  \vspace{0mm}
\end{figure}

In this study, the SVM was implemented within MATLAB's SVM solver \textit{fitcsvm}, with default RBF parameters.
The hyperparameters $n_{init}$ and $n_{learn}$ were set to 100 and 30 respectively by tuning for time and sample efficiency for the 3 test functions of Eq. \eqref{eq:testFx}.

This is illustrated using the Beale test function $h(x,y)$ in Fig. \ref{IM:Beale}, where the true boundary $\partial S$ is shown by the black contour.
ALB first randomly samples $n_{init}$ points in the domain $\X$, which are then labeled and used for training the SVM.
In this example, the SVM approximates $\partial S$ as the three disjoint boundaries shown in Fig. \ref{im:Beale1}, which is markedly different from the true boundary.

For the next iteration, sufficiently well-spaced points, marked by the yellow squares in Fig. \ref{im:Beale1}, are selected for querying by $h(x,y)$.
These points are selected by \textit{k-means} clustering on the predicted boundary to define $n_{learn}$ clusters, after which, the medoid of each cluster, i.e., the predicted boundary point closest to the centroid of that cluster is selected for querying.
These newly labeled points are added to the training set, and the process is repeated.
As seen in Fig. \ref{im:Beale5}, after 15 iterations, the predicted boundary better approximates $\partial S$.
For comparative evaluation, the sample budget of the ALB algorithm is limited to the number of samples used by the EDGE algorithm.
The resulting ALB-localized boundary shown in Fig. \ref{im:BealeEnd} does not capture the nonlinear regions of $\partial S$ as closely as the EDGE boundary shown in Fig. \ref{im:BealeEDGE}. 

\begin{figure}[t]
    \centering
      \vspace{0mm}
\subfloat[ALB Sample Efficiency \label{im:sampleConv} ]{%
       \includegraphics[trim =70mm 110mm 75mm 113mm, clip, width=0.49\linewidth]{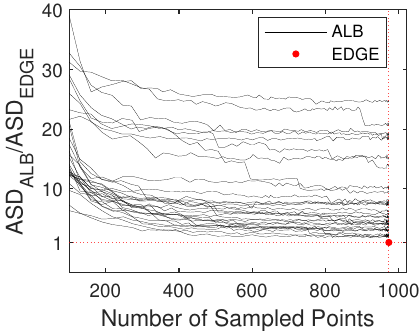}}
       \hspace{0mm}
\subfloat[ALB Time Efficiency \label{im:timeConv}]{%
       \includegraphics[trim =70mm 110mm 77mm 113mm, clip, width=0.49\linewidth]{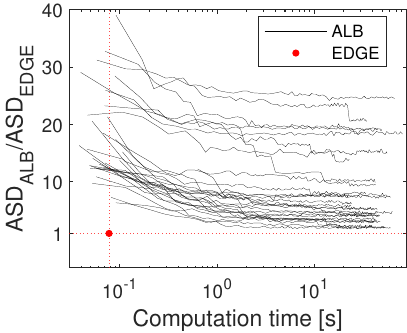}}
  \caption{Rosenbrock Test Function ($\varepsilon = 0.05$)}
  \label{IM:Rosenbrock} 
  \vspace{0mm}
\end{figure}

In this paper, all experiments were conducted in a Matlab R2023b environment using an AMD Ryzen 5 5600X CPU clocked at 3.7 GHz, equipped with 128 GB RAM.
To account for stochasticity associated with initialization through random points, the ALB algorithm is repeated 30 times.
The convergence of ALB's accuracy as a ratio of its ASD with that of EDGE is plotted in Fig. \ref{IM:Rosenbrock} for the Rosenbrock test function.
For $\varepsilon$ values of $0.1$ and $0.05$, the ASD accuracy, sample efficiency and computation time are noted for the 3 test functions and 3 algorithms in Table \ref{tab:perfComp}.

While the ALB algorithm reduces its approximation error, i.e., ASD, by iteratively refining its localization of $\partial S$, its performance is influenced by the $n_{init}$ initialization points, as evident in the significant stochasticity seen in Fig. \ref{IM:Rosenbrock} and listed in Table \ref{tab:perfComp}.
Contrarily, the EDGE algorithm showed no stochasticity, had lower ASD values across ALB repetitions, test functions, and $\varepsilon$ values. 
The ALB method also required higher computation time due to SVM's time complexity of $\mathcal{O}(m^3)$ due to kernel matrix computations. 

The $\varepsilon$-neighborhood guarantee is absent in the ALB, but is provided by both the EDGE algorithm and the na\"ive grid-search method.
As seen in Table \ref{tab:perfComp} however, the sample efficiency of the EDGE algorithm is vastly superior to the grid-search method while providing similar ASD values. 
The grid-search method also scales poorly because the sample count over a rectangular domain of sides $a,b$ is given by $ab/\varepsilon^2$, and thus, for the same $\partial S$ perimeter $L$, if the lengths $a,b$ are doubled or $\varepsilon$ is halved, the number of na\"ive approach queries is quadrupled.
\textcolor{black}{
In contrast, the EDGE sample count scales with $L/\varepsilon$.
}
\section{Conclusion}

The proposed EDGE algorithm leverages the intermediate value theorem to approximate the decision boundary using greedy geometric operations.
By doing so, EDGE outperforms active learning approaches in both boundary approximation accuracy and computational efficiency.
Unlike active learning, EDGE guarantees that its estimated boundary is within the $\varepsilon$-neighborhood of the true boundary.
Moreover, EDGE does not require hyperparameter tuning, making it applicable to diverse black-box systems.

Future work will extend the current binary framework to multiclass problems by decomposing the problem into multiple binary classification subproblems.  
A natural extension to EDGE is to utilize its efficiency in high dimensional spaces where boundary coverage and sample efficiency are key to a successful implementation.
The framework will also be applied to black box systems with system or measurement noise, using the parameter $\varepsilon$ to accommodate the soft transition zone between classification regions. 

Further, in the context of SVMs, the labeled boundary points identified by EDGE serve as support vectors that heavily influence classification performance.  
Since SVMs rely on identifying samples close to the decision boundary, training an SVM using the labeled point set generated by EDGE will significantly reduce stochastic behavior and enhance sample efficiency by enriching the training set with relevant boundary information.


\vspace{-1mm}

\bibliography{ifacconf}             

\end{document}